\documentstyle[11pt,aaspp4]{article}
\catcode`@=11
\def\lsim{\mathrel{\mathpalette\@versim<}}    
\def\gsim{\mathrel{\mathpalette\@versim>}}    
\def\@versim#1#2{\lower0.2ex\vbox{\baselineskip\z@skip\lineskip\z@skip
  \lineskiplimit\z@\ialign{$\m@th#1\hfil##\hfil$\crcr#2\crcr\sim\crcr}}}
\catcode`@=12

\received{2000 August 11}
\accepted{2001 February 5, for {\it Ap. J.}}

\lefthead{Takahashi et al.}
\righthead{The fastest millisecond pulsar PSR B1937+21}

\begin{document}

\title{
Pulsed X-ray Emission from The Fastest Millisecond Pulsar
PSR B1937+21 with ASCA}

\author{
Motoki Takahashi\altaffilmark{1,8},
Shinpei Shibata\altaffilmark{1}, 
Ken'ichi Torii\altaffilmark{2},
Yoshitaka Saito\altaffilmark{3},
Nobuyuki Kawai\altaffilmark{4,2},
Masaharu Hirayama\altaffilmark{5},
Tadayasu Dotani\altaffilmark{3},
Shuichi Gunji\altaffilmark{1}, 
Hirohisa Sakurai\altaffilmark{1},
Ingrid H. Stairs\altaffilmark{6,9}, and
Richard N. Manchester\altaffilmark{7}
}

\altaffiltext{1}{Department of Physics Yamagata University, Kojirakawa, Yamagata 990-8560, Japan%
; shibata@sci.kj.yamagata-u.ac.jp
}
\altaffiltext{2}{National Space Development Agency of Japan, Tsukuba Space Center, Space Utilization Research Program, 2-1-1 Sengen, Tsukuba, Ibaraki 305-8505, Japan}
\altaffiltext{3}{The Institute of Space and Astronautical Science (ISAS), Sagamihara, Kanagawa 229-8510, Japan}
\altaffiltext{4}{The Institute of Physical and Chemical Research (RIKEN), Wako, Saitama 351-01, Japan}
\altaffiltext{5}{Santa Cruz Institute for Particle Physics, 
University of California, Santa Cruz, CA 95064}
\altaffiltext{6}{University of Manchester, Jodrell Bank Observatory,
Macclesfield, Cheshire, SK11 9DL, UK}
\altaffiltext{7}{Australia Telescope National Facility, CSIRO, Radiophysics,
P.O. Box 76, Epping, NSW 1710, Australia}
\altaffiltext{8}{present address: Kunori-Gakuen High School, 1-1-72, Monto-machi, Yonezawa, 992-0039, Japan}
\altaffiltext{9}{present address: NRAO, P.O. Box 2, Green Bank, WV 24944}

\begin{abstract}

We have detected pulsed X-ray emission
from the fastest millisecond pulsar known, PSR B1937$+$21 ($P=$1.558 msec), 
with ASCA.
The pulsar is detected as a point source above $\sim 1.7$ keV,
with no indication of nebulosity.
The source flux in the 2--10 keV band is found to be
$f = (3.7\pm 0.6) \times 10^{-13}$ erg s$^{-1}$ cm$^{-2}$,
which implies an isotropic luminosity of
$L_{\rm x} = 4 \pi D^2 f$
$\sim (5.7\pm 1.0) \times 10^{32} ~(D/3.6 \; {\rm kpc})^2$ erg s$^{-1}$, 
where $D$ is the distance,
and an X-ray efficiency of $\sim 5 \times 10^{-4}$ relative 
to the spin-down power of the pulsar.

The pulsation is found at the period predicted by the radio ephemeris
with a very narrow primary peak, the width of which is about
1/16 phase ($\sim 100 \; \mu$s),  near the time resolution limit ($61 \; \mu$s)
of the observation.
The instantaneous flux in the primary peak (1/16 phase interval) is
found to be 
($4.0\pm 0.8) \times 10^{-12}$ erg s$^{-1}$ cm$^{-2}$.
Although there is an indication for the secondary peak,
we consider its statistical significance too low to claim a definite detection.

The narrow pulse profile and the detection in the 2--10 keV band
imply that the X-ray emission is caused by the magnetospheric 
particle acceleration.
Comparison of X-ray and radio arrival times of pulses
indicates, within the timing errors, 
that the X-ray pulse is coincident with the radio interpulse.

\end{abstract}


\section{Introduction}

There are two proposed emission regions in
the pulsar magnetosphere, the polar cap near the surface of the neutron
star and 
the outer gap near the light cylinder.
The magnetic field strength of the polar cap is 
typically $10^{12}$ G for the {\em ordinary pulsars},
while it is typically $10^8 - 10^9$ G, three or four orders of
magnitude smaller, for the {\em millisecond pulsars}.
The field strength in the outer gaps near the light cylinder is about $10^6$ G
for very young ordinary pulsars such as the Crab, 
while it decreases considerably with spin-down because 
the light cylinder becomes large:
for a period of $P=0.1$ s, 
the light cylinder field $B_{\rm L}$ becomes typically $2 \times 10^4$ G.
For instance, for Vela, $P=0.089$ s and $B_{\rm L} = 4.4 \times 10^4$ G.
The field strength at the light cylinder of the millisecond pulsars
can exceed the value of the Crab pulsar because of small light
cylinder radii.
Therefore, if high energy emission from the pulsar magnetosphere
is affected by the field strength, 
comparison of the two pulsar populations in high energy radiation
will, plausibly, elucidate the emission and acceleration mechanism.
In particular, detections of the millisecond pulsars in X-rays 
with ROSAT, ASCA, RXTE, Beppo SAX and recently Chandra and XMM Newton
are expanding our knowledge of the X-ray emission from the
millisecond pulsars.

The fastest millisecond pulsar known is PSR B1937$+$21 (\cite{bak82}),
which was previously observed with the ROSAT HRI, 
giving an upper limit on the flux 
(\cite{ver96}).
In this paper, 
we report the detection of narrow pulses from this pulsar
for energies above $\sim 1.7$ keV in a $100$ ks ASCA observation.
We give the results of
the spatial, temporal and spectral analyses. 
A comparison with other millisecond pulsars is also given.

\section{Observation}

We observed PSR B1937+21 with ASCA (\cite{tih94}) on 15--17 November 
1997 (50767.959722 -- 50769.084027 MJD) for a duration of 97,140 s. 
The Gas Imaging Spectrometer (GIS; \cite{mak96}, \cite{oha96})
was operated at the highest time resolution, 61$\mu$s.
In this operation mode, the highest time resolution is obtained 
at the expense of spatial resolution:
the GIS field of view consists of 64 $\times$ 64 pixels of  
$\sim 0^{\prime}.98 \times 0^{\prime}.98$ each.
The Solid-state Imaging Spectrometer (SIS; \cite{bur91}, \cite{bur94}) 
was operated in 1-CCD FAINT mode. 
We screened the event data by the standard selection criteria. 
The effective exposure time is 57.2 ks for the SIS and 60.0 ks for the GIS,
 including 38,209 s exposure with the highest time resolution.

\section{Spatial Analysis}

We present the X-ray contour maps in 
the GIS soft band (0.7--2 keV) and the GIS hard band (2--10 keV)
in Figure~1, and contour maps 
in the SIS soft band (0.5--2 keV) and in the SIS hard band (2--10 keV)
in Figure~2.
In the hard band images of both the GIS and the SIS,
we clearly detect a source spatially coincident 
with the radio position within an accuracy of 40$^{\prime\prime}$ 
(90 $\%$ confidence taking into account pointing uncertainties; 
see Gotthelf 1996). 

\begin{figure}
\plottwo{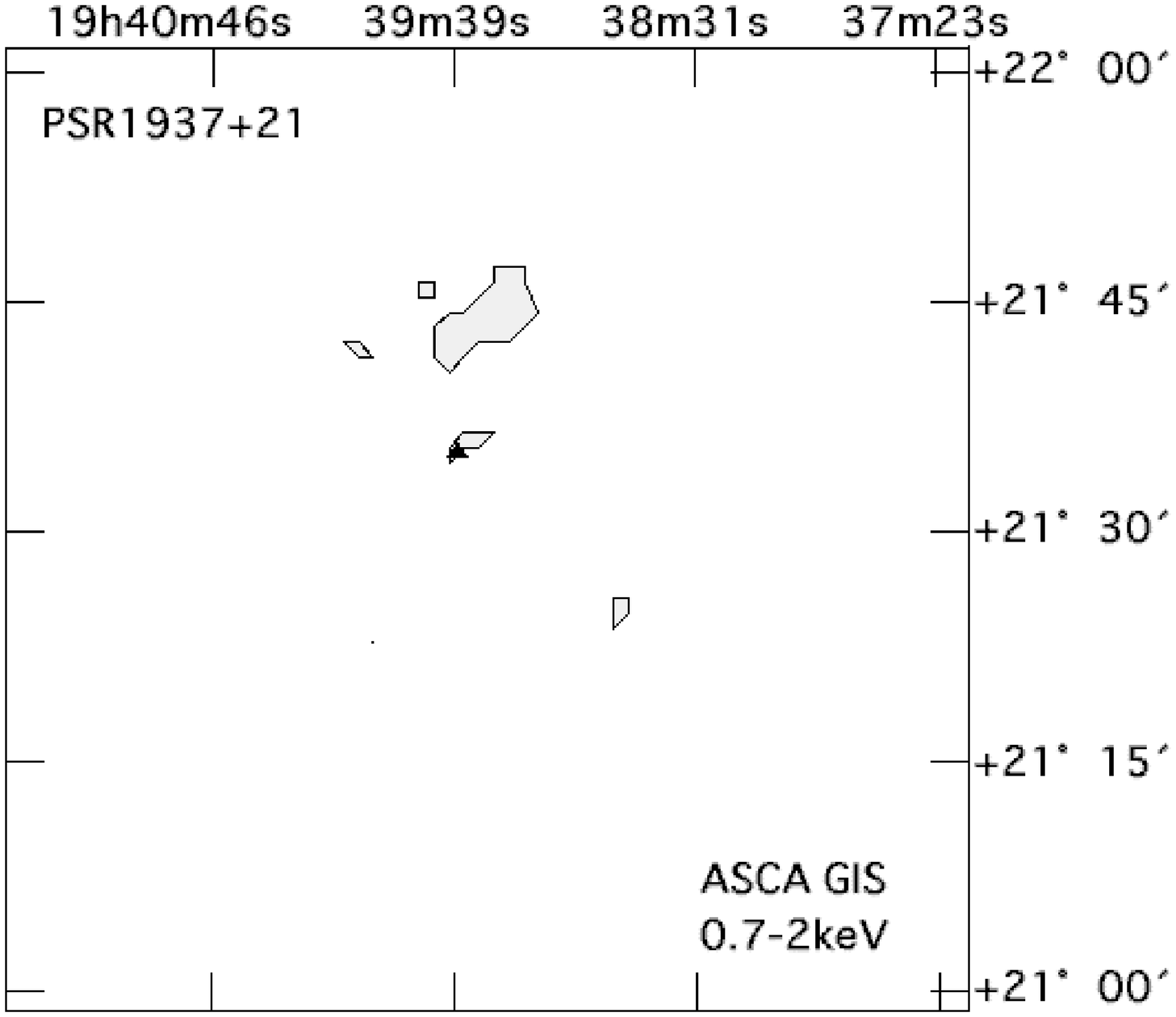}{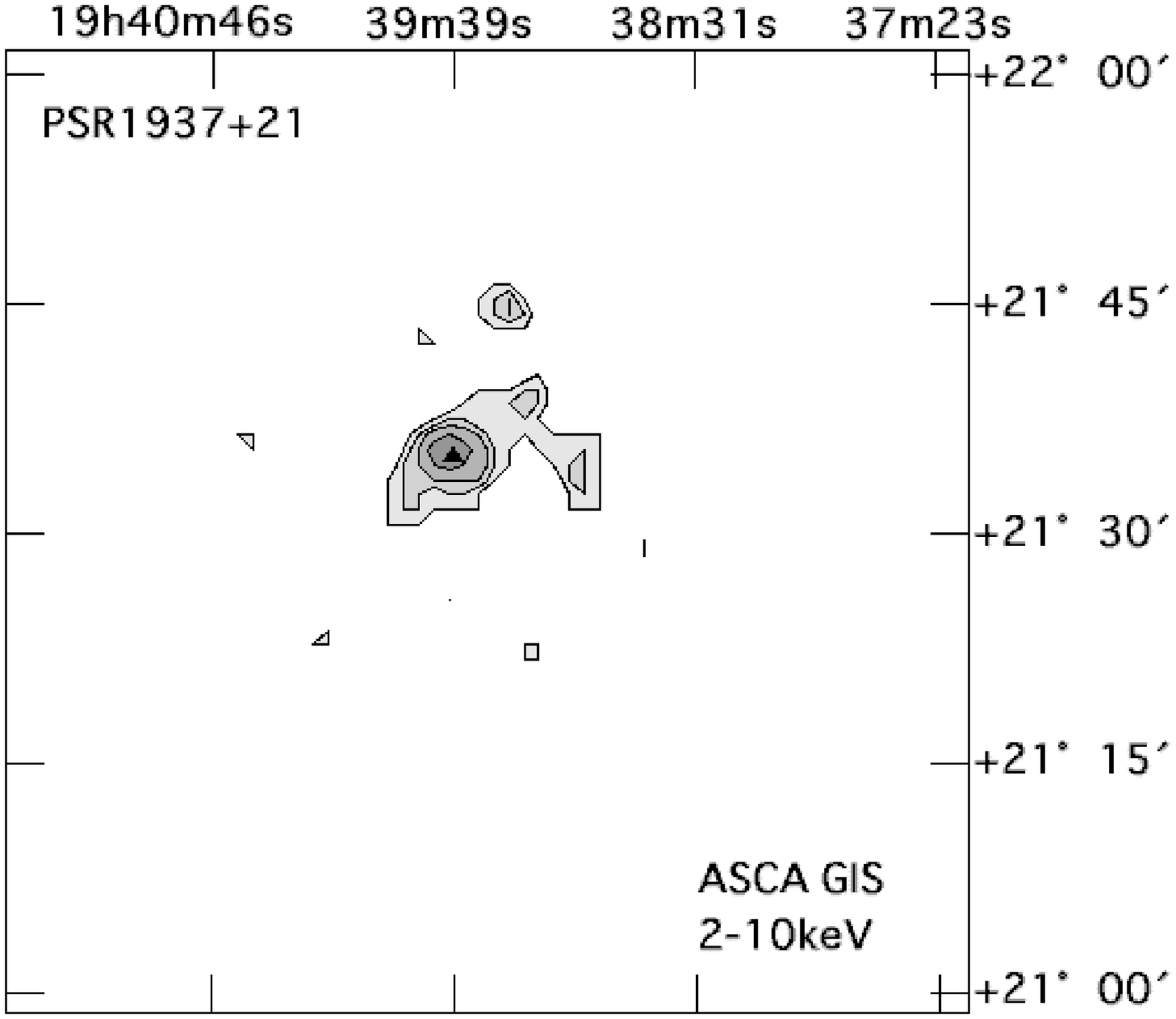}
\caption{X-ray contour maps of PSR~B1937$+$21 obtained with GIS
in the 0.7--2 keV band ({\it left}) and in the 2--10 keV band ({\it right}).
The contour for the soft band ({\it left}) is 5 counts per pixel, 
corresponding to $4.1 \sigma$, and
the contours for the hard band ({\it right}) are 8, 9, 10 and 12 
counts per pixel,
corresponding to $3.0 \sigma$, $5.1 \sigma$, $7.2 \sigma$ and $13 \sigma$,
respectively.
The filled triangles in the contour maps indicate the radio position of the pulsar.
\label{fig1}}
\end{figure}

\begin{figure}
\plottwo{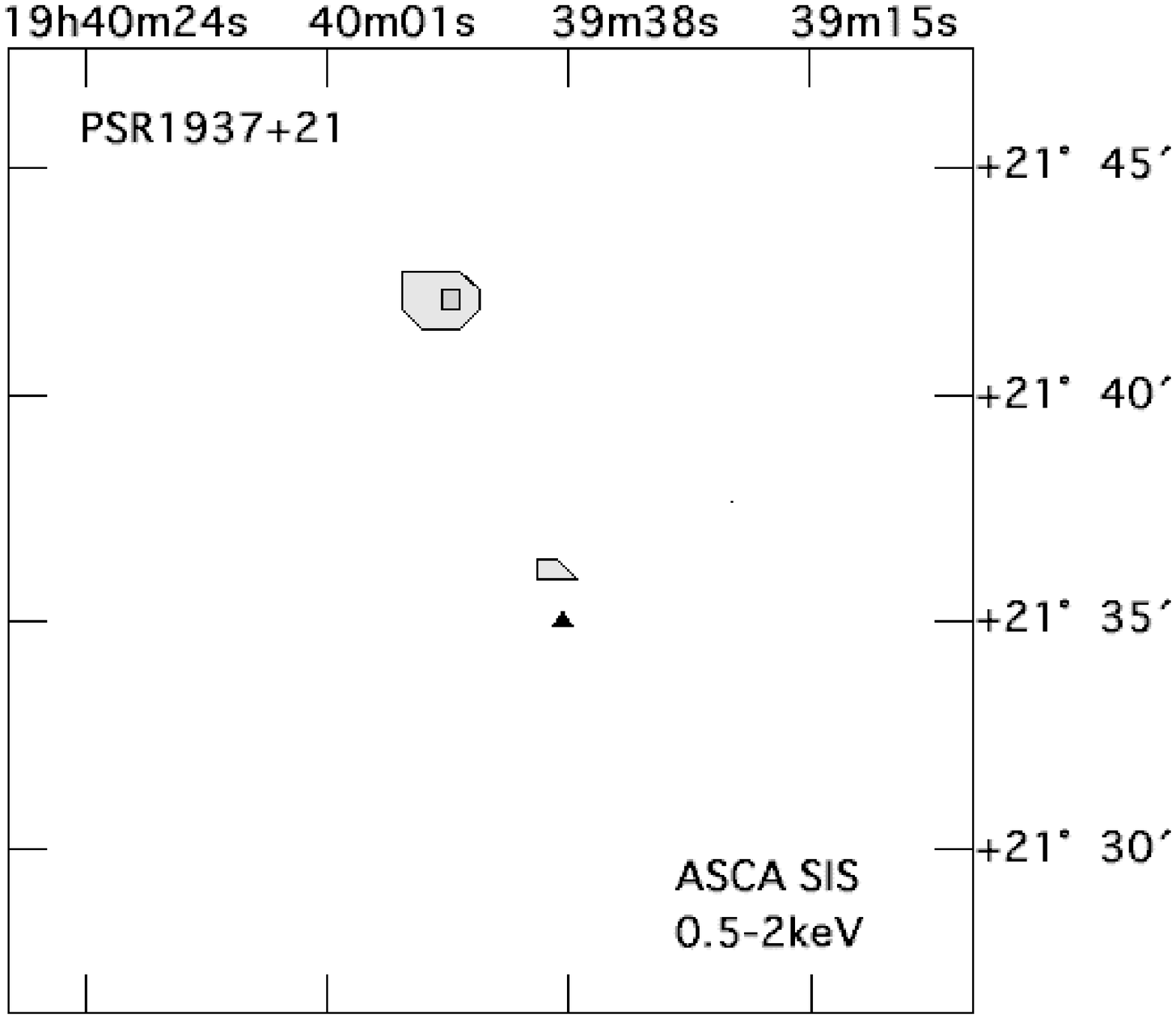}{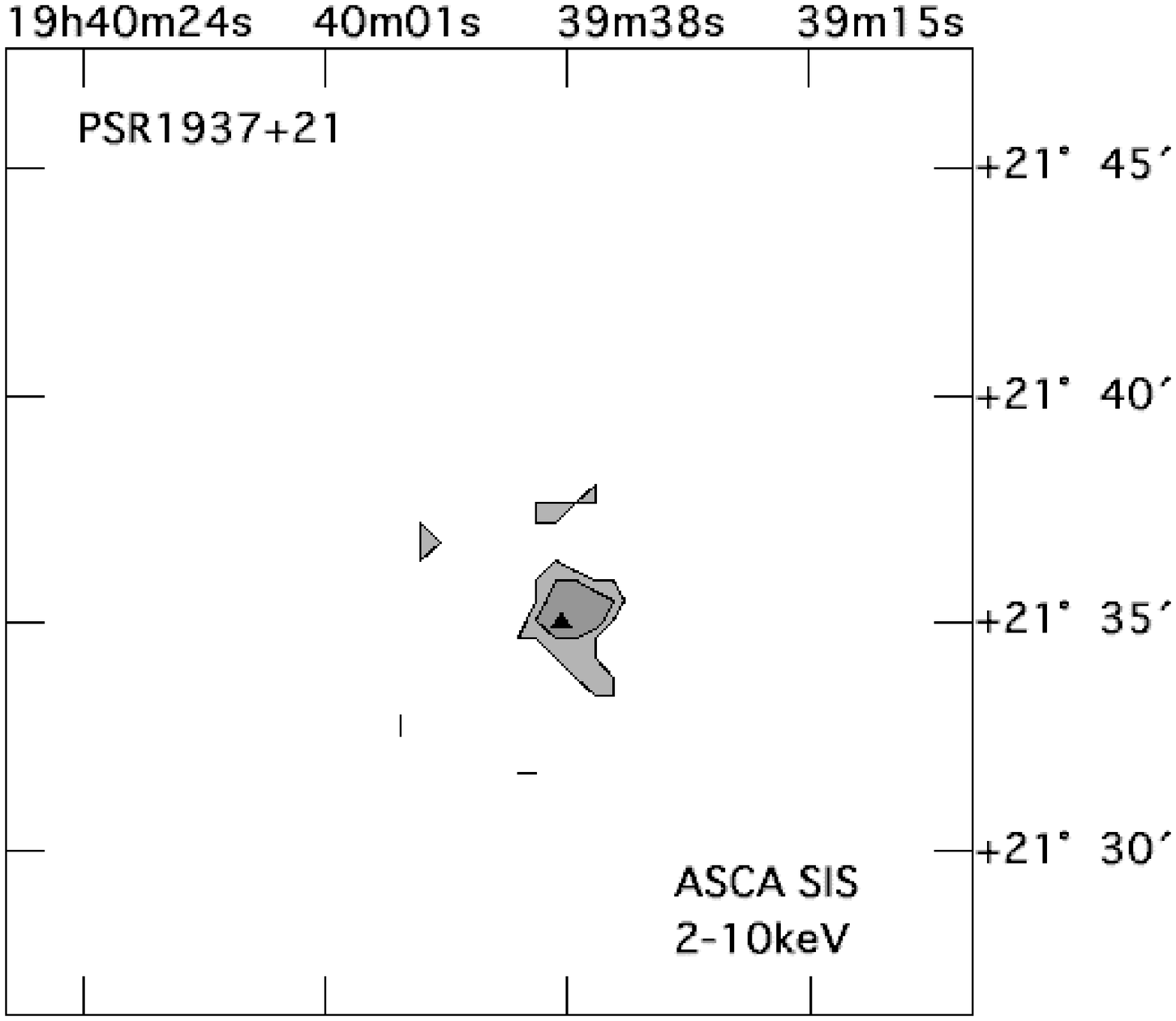}
\caption{X-ray contour map of PSR~B1937+21 obtained with SIS
in the 0.5--2 keV band ({\it left}) and in the 2--10 keV band ({\it right}).
The contours for the soft band ({\it left}) are 2 and 3 counts per pixel,
corresponding to $4.5 \sigma$, and $10 \sigma$, respectively, and
the contours for the hard band ({\it right}) are 3 and 4 counts per pixel,
corresponding to $5.2 \sigma$ and $9.9 \sigma$, respectively.
The filled triangles in the contour maps indicate the radio position of the pulsar.
\label{fig2}}
\end{figure}

\placefigure{fig1}
\placefigure{fig2}

In order to estimate the background count rate,
we take the background region as follows.
The GIS background region is a circular region with radius of 20$^\prime$ 
centered on the detector center with subtraction of a circular region 
with radius of 6$^\prime$ centered on the pulsar and an elliptical region 
with a semi-major axis of 7$^\prime$ and a semi-minor axis 
of 4$^\prime$ centered 
on the soft source seen north of the pulsar (Fig.~1; left panels). 
Similarly, the SIS background region is the whole chip from which we removed a 
circular region with radius of 3$^\prime$ centered on the pulsar 
and an elliptical region with a semi-major axis of 2$^\prime$ and a 
semi-minor axis of 1$^\prime$ centered on the soft source north 
of the pulsar. 

If we define the source region as a circle with radius 
3$^\prime$ for the GIS and 1.5$^\prime$ for the SIS, 
centered on the radio position, 
the source significance, which is estimated by the source flux
divided by the background fluctuation, 
is found to be $\sim$ 15$\sigma$ in the GIS hard 
band (2-10 keV) and $\sim$ 5$\sigma$ in the SIS hard band (2-10 keV).
(These radii of the source regions were found to maximize the
source significance; see the last part of this section.)
On the other hand, we could not detect the source in the soft 
band images.
Using PIMMS\footnote{%
http://heasarc.gsfc.nasa.gov/Tools/w3pimms.html}
we find that the present result is consistent 
with the ROSAT non-detection in the
0.1--2.4 keV energy range 
subject to an inferred interstellar absorption which 
corresponds to a hydrogen column density of
$(0.28 - 2.2) \times 10^{21}$ cm$^{-2}$
(\cite{ver96}) and to an assumption of
a power law spectrum with index of $\sim 1$
as suggested by the spectral analysis given below.

We compared the spatial distribution of the events with 
the point spread function, finding that the source is 
consistent with a point source. 
For this analysis, 
the GIS image is not suitable, but we used 
the SIS image in the 0.7--10 keV band. 
Considering the pointing instability, 
we find that a possible spatial extent must be smaller than 
$10^{\prime\prime}$.

For the timing and spectral analyses,
we optimized the radius of the source 
region so that the signal-to-noise ratio was maximized:
we searched for the radius for which the source flux divided by the
background flux was maximized. 
In the same way, 
for the timing analysis, the optimized energy window was also obtained 
for the GIS data for which the highest time resolution is available. 
As a result, 
the optimized radius was found to be 3.0$^\prime$ for the GIS data and 
1.5$^\prime$ for the SIS data; 
the optimized energy window was 1.7--6.5 keV for the 
timing analysis.

\section{Timing Analysis}

In the previous section we obtained the optimized 
spatial and energy regions, i.e.,  
a circular region with the radius 3.0$^\prime$ and an energy
window of 1.7--6.5 keV. 
Applying these optimized spatial and energy regions, 
we collected 227 photons (99 for GIS2 and 128 for GIS3).
After converting the arrival times on ASCA to those at the solar system 
barycenter using TIMECONV\footnote{
Earlier versions are not suitable the fastest pulsars, 
because leap seconds are ignored 
in the calculation of the position of the satellite.\\ 
See in detail http://heasarc.gsfc.nasa.gov/docs/asca/newtimeconv.html
} version 1.53, 
we searched for a periodicity by 
folding the photon arrival times into 32 bins with trial periods around
the radio period expected from the ephemeris given in Table~1. 
The $\chi^2$ values calculated relative to a flat distribution are 
plotted against the trial periods to make a periodogram 
(Fig.~3). 
We found a peak of $\chi^2$ = 212 (dof $=31$) at 
$P_{\rm ASCA}$ = 1.557806498(9) ms, which coincides with the predicted 
$P_{\rm radio}$ = 1.55780649848632017 ms 
based on the radio ephemerides (Table~1) with an accuracy higher than
$\sim 10^{-11}$ s, which is expected from the duration of 
the X-ray observation. 
The peak significance corresponds to 11.15 $\sigma$
(chance probability of $7.4 \times 10^{-29}$).
We also applied the bin free $z_n^2$ test (\cite{bch83}) using 
$n$-harmonics with sevearl $n$'s and
confirmed the result.
Figure~3 illustrates the result of the period search.

\begin{figure}
\plotone{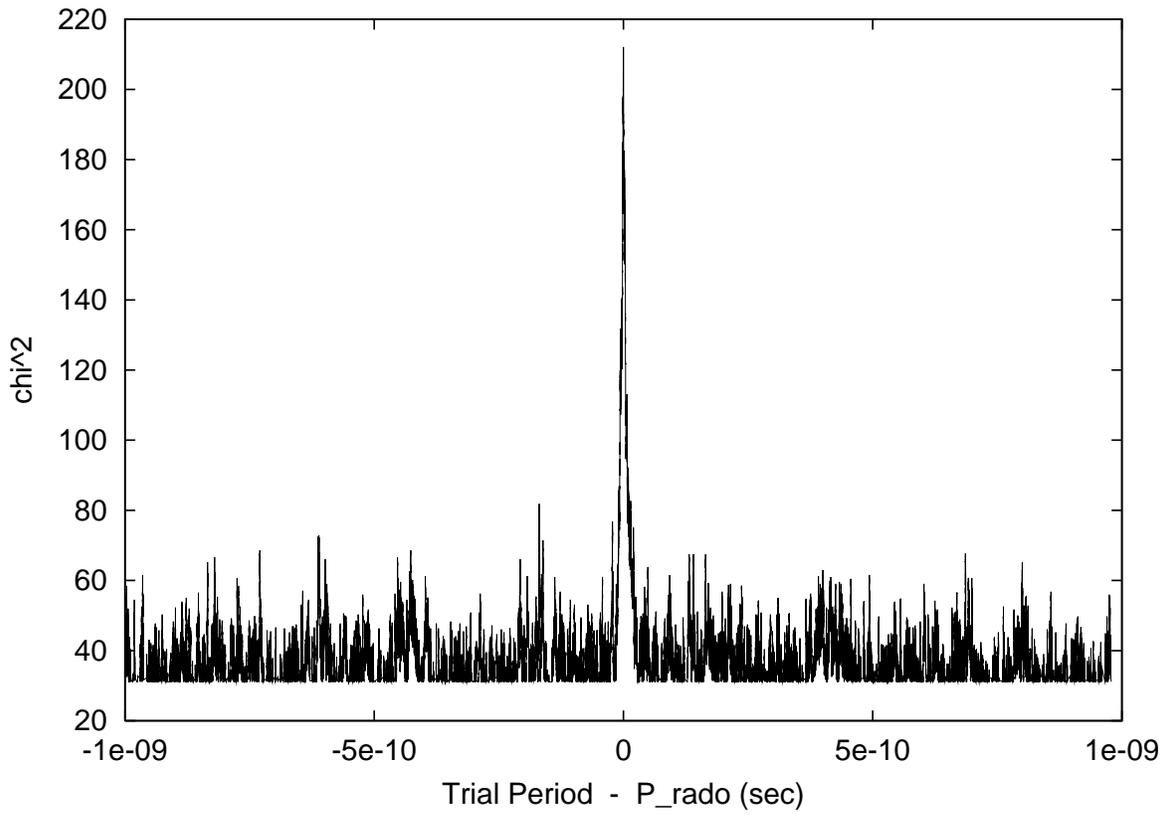} 
\caption{%
The $\chi^2$ significances are
plotted as a function of trial period. 
The zero point corresponds to the
expected period derived from the radio ephemeris (Table 1)
\label{fig3}}
\end{figure}

\placefigure{fig3}

We subsequently folded the data using the predicted 
period $P_{\rm radio}$, finding
the sharply peaked profile shown in the upper panel of 
Figure~4. It is very difficult to 
estimate the pulse width accurately because the width is close to the GIS 
time resolution (61 $\mu$s). By trying various numbers of phase 
bins, we estimated it to be 1/16 (=0.0625) phase ($\sim$ 100 $\mu$s). 
We examined  the energy dependence of the pulse profile,  
comparing profiles for the softer and harder energy bands, 
2--5 keV and 5--10 keV. 
No significant difference was found. 

\begin{figure}
\begin{center}
\mbox{\epsfysize=9cm \epsfbox{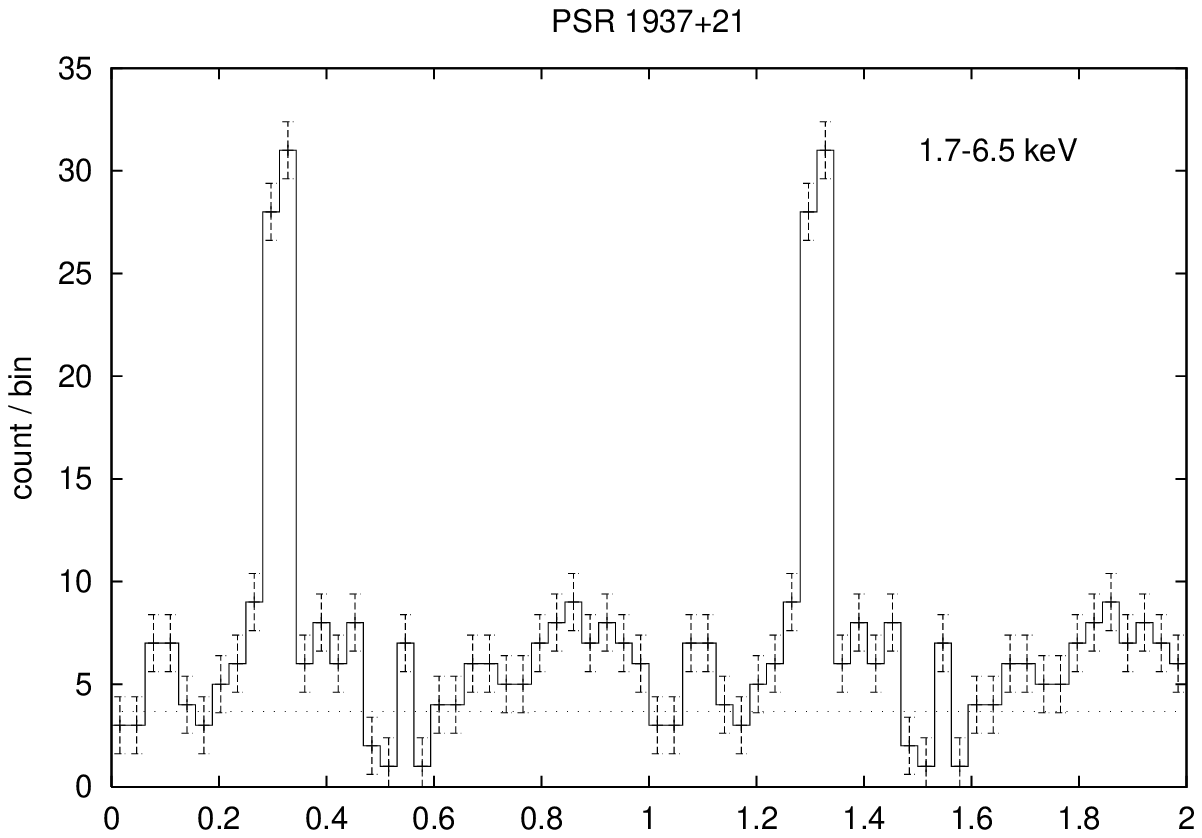}} \\
\mbox{\epsfysize=9cm \epsfbox{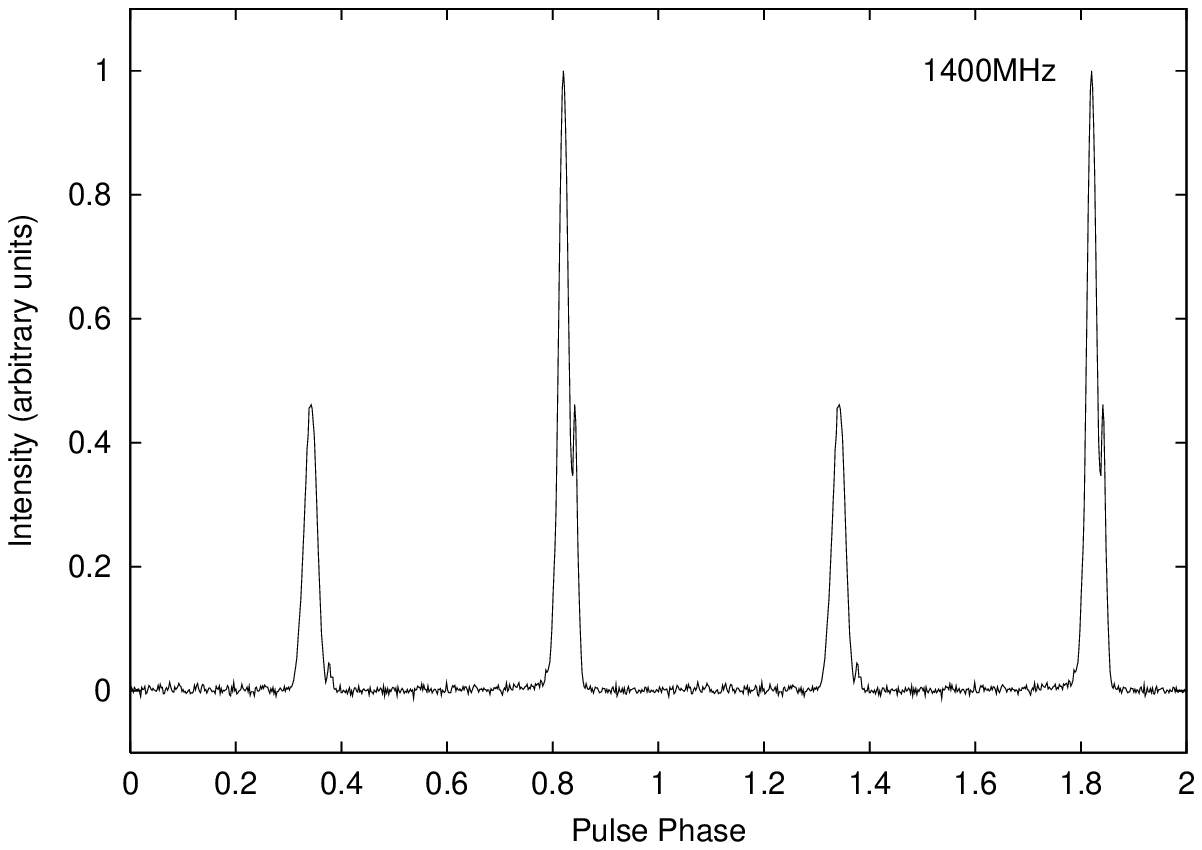}}
\caption{{\it Upper panel}: 
Pulse profile of PSR B1937+21 in the GIS 1.7 - 6.5 keV 
band. Two phase cycles are shown for clarity.
The dashed line indicates the background level, 3.67 counts/bin, 
for a division in 32 phase bins. The bin size corresponds
to $\sim 49 \mu$s.
{\it Lower panel}:
The radio pulse profile at 1.4 GHz. We find that the X-ray pulse is
coincident with the radio interpulse.
The timing error for the alignment is about 15~$\mu$s in the radio 
data and 100~$\mu$s (0.064 phase) in the X-ray data:  
the timing error is roughly two phase bins in the upper panel. 
\label{fig4}}
\end{center}
\end{figure}

\placefigure{fig4}

In the X-ray pulse profile (Fig.~4; upper panel), where we used 32 phase bins,
the background level is 3.67 counts/bin indicated by the dashed line. 
The excess above the background level for the `on-pulse' defined by 
the phase interval 8.5/32 -- 10.5/32 (0.266 -- 0.328) is 51.66 counts 
(19 $\sigma$), 
and the excess for the `off-pulse' is 1.93 counts/bin,
which is considered to be the DC component. 
Therefore, 
out of 51.66 counts in `on-pulse' phase interval,
47.80 counts belong to the pulsed radiation, from which
we find a pulsed fraction of $\sim$ 44$\%$.
One may notice a possible secondary peak at the phase interval 
24.5/32 -- 30.5/32 (0.766--0.953). 
The excess counts for this phase interval above the background 
is 23.98 counts; i.e., $\sim$ 5 $\sigma$ above the 
background over these 6 bins. If we regard this excess 
as the pulsed radiation, then the pulsed fraction 
is calculated to be $\sim$ 59$\%$. 
Although the presence of a secondary peak is common in high energy
emission from pulsars, 
in the present observation
the limited number of photons does not allow us to 
determine the count rates for both the secondary pulse component and 
the DC component simultaneously.
Therefore, the presence of the secondary pulse is only tentative.

Since the narrow X-ray pulses were tracked through the observation, 
a previous estimate (\cite{sai97}) of the ASCA clock stability may 
be improved to   
$|\delta f|/f \le 8 \times 10^{-8}$ for the duration of $\sim$100 ks.   

We have investigated the alignment of the X-ray pulse peak relative to the
radio pulse profile.  We used 1400 MHz and 600 MHz observations from the
76-m Lovell Telescope at Jodrell Bank Observatory performed between 1997 Sept.
19 and 1997 Nov. 3, and 430 MHz observations from the 300-m Arecibo
telescope performed on 1997 Nov. 22.  
The radio pulse profiles are very similar at all frequencies used, 
with the same relative phases of the main pulse and interpulse 
(\cite{krm99}). 
We aligned these radio data to the same fiducial point, 
then used the standard pulsar timing program {\sc tempo}
(http://pulsar.princeton.edu/tempo) to obtain a good local value of the
dispersion measure and the ephemeris given in Table 1.  
We subsequently included the X-ray data in the fit, 
counting the geocentric arrival time
of each X-ray photon within the pulse peak as equivalent to a radio pulse
arrival time at infinite frequency.  We find that the X-ray peak falls
$767\, \pm \,15\, \pm \,100 \; \mu$s after the peak of the radio profile main pulse. 
The latter part of the error comes from the systematic error in 
absolute time calibration to the ASCA onboard clock.
The radio profile interpulse occurs approximately 800\,$\mu$s after the main
pulse, so it is plausible, within the timing errors, that the X-ray pulse
is coincident with the radio interpulse.
In Figure~4 the radio profile is aligned with the X-ray profile.

\section{Spectral Analysis}

Spectral analysis was performed on the GIS data for which the source
region yielded $\sim$ 135 photons after background subtraction.  We
also tried spectral fits for the data restricted to the `on-pulse'
phase. Taking advantage of the phase selection, which enabled us to
have a better signal-to-noise ratio, we used a larger aperture of
radius 6$^{\prime}$ for the source region; the background region was
the same as before but only the `off-pulse' data were used. In this
fashion we collected $\sim$ 84 photons.

Unfortunately, the statistics were not sufficient for reliable spectral fits. 
However, for interstellar hydrogen column densities between
$N_{\rm H}$ = 2.8 $\times$ 10$^{20}$ cm$^{-2}$ 
(see the ROSAT HRI observation; Verbunt et al. 1996) and 
$N_{\rm H}$ = 1.4 $\times$ 10$^{22}$ cm$^{-2}$ 
(a typical Galactic hydrogen column density in the direction of the source), 
we could fit the spectra by power law models 
(typical reduced $\chi^{2}_{\rm r}$
 $\sim$ 0.8, d.o.f. = 28). The source flux was 
found to be (3.7$\pm$ 0.6) $\times$ 10$^{-13}$ 
erg s$^{-1}$ cm$^{-2}$ in the 2 -- 10 keV band 
with a photon index\footnote{%
The photon index $\alpha$ is defined such that the photon flux
is given by $K \epsilon^{- \alpha}$, where 
$\epsilon$ is the photon energy and $K$ is the normalization.
We used a power law with pegged normalization
(see the XSPEC reference; \cite{arn96}).}
of 0.8$\pm$0.6. 
On-pulse data were also fitted by a power law model and
we found an instantaneous pulsed flux of
$(4.0 \pm 0.8) \times 10^{-12}$ erg s$^{-1}$ cm$^{-2}$ in the 2-10 keV band.

The source is detected in the hard band (2--10 keV) and shows a very
narrow pulse. These facts imply that the X-ray pulsed emission has
a magnetospheric non-thermal origin.

The observed flux indicates an (isotropic)  X-ray luminosity of 
$L_{\rm x} = 4 \pi D^2 f$ $\approx (5.7\pm 1.0) \times 10^{32} 
(D/3.6 \; {\rm kpc})^2$ erg s$^{-1}$, which corresponds to 
5 $\times 10^{-4}$ of the spin-down power of the pulsar.
The distance estimate of 3.6 kpc is based on the dispersion measure
(\cite{tcx93}) and is near the lower bound
on the distance obtained by annual parallax (\cite{ktr94}).
If the power law spectrum extends to higher energies with
a similar photon index $\sim 1$, the X-ray luminosity
would become as large as the spin-down power at about  a few MeV.
Therefore, there must be a turn off below a few MeV,
consistent with the EGRET upper bound (\cite{fie95}).

\section{Summary and Discussion}

Our detection of pulsed emission in the 1.7--6.5 keV band
from PSR~1937$+$21  reveals that the pulse is 
very narrow with a width of $\sim 100 \; \mu$s ($\sim 0.06$ in phase),
indicating a non-thermal magnetospheric origin.
There is an enhancement indicating a secondary peak, but its
statistical significance is too low to be conclusive.
Due to this uncertainty, we can only estimate the pulsed fraction,
placing it at $\sim 40-60$~\%.
The pulse alignment shows, within the timing errors, 
that the X-ray pulse is coincident with the radio interpulse.
The 2--10 keV band flux (pulse $+$ DC component) is  
$f = (3.7\pm 0.6) \times 10^{-13}$erg s$^{-1}$ cm$^{-2}$,
which implies an isotropic luminosity of
$L_{\rm x} = 4 \pi D^2 f$
$\sim (5.7\pm 1.0) \times 10^{32} ~(D/3.6 \; {\rm kpc})^2$ erg s$^{-1}$,
where $D$ is the distance.
The X-ray efficiency is $\sim 5 \times 10^{-4}$ 
of the spin-down power of the pulsar.

Based on ROSAT observation of the rotation powered pulsars,
Becker and Tr\"{u}mper (1997) examined the correlation between
the 0.1--2.4 keV band luminosity and the spin-down power $\dot{E}$.
They showed that regardless the pulsar population, i.e., ordinary
or millisecond pulsars, the X-ray luminosity follows
an empirical law of 
$ L_{\rm x}(0.1-2.4 \; {\rm keV}) = 10^{-3} \dot{E}$.

Saito (1997) searched ASCA data for a similar correlation in
the 2-10 keV band. In Figure~5, we re-examine the $L_{\rm x}$-$\dot{E}$
correlation in the 2--10 keV band,
adding the results of this study and
recent observation of PSR~J2124--3358 (\cite{sak99}).
It is confirmed that the correlation is common to both populations 
and follows $L_{\rm x}(2-10 \; {\rm keV}) \propto \dot{E}^{1.5}$.

\begin{figure}
\begin{center}
\plotone{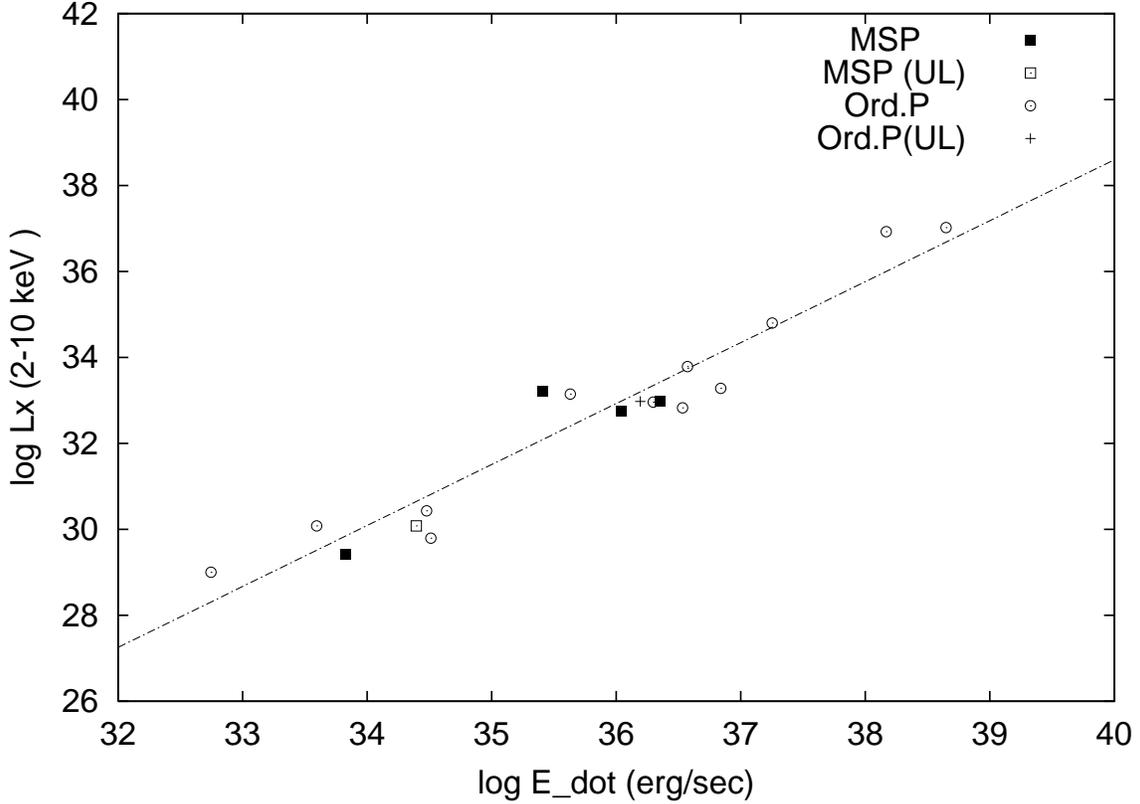} 
\caption{Pulsar luminosity in the 2--10 keV band versus the rotation power of the
pulsars. The data are taken from Sakurai et al. (1999) for PSR~J2124-3358, 
Pivovaroff, Kaspi \& Gotthelf (2000) for PSR~B1046--58 and PSR~B1610--50, 
Mineo et al. (2000) for PSR~J0218+4232,
Saito (1997) for others,
and  the present work for PSR~1937$+$21.
The filled boxes indicate millisecond pulsars, while the open box indicates
an upper limit (PSR~J0437--4715).
The open circles indicate ordinary pulsars, and the cross indicates an
upper limit (PSR~B1610--50).
The regression line is given by 
$\log L_{\rm x} = 1.45 (\pm 0.09) \log \dot{E} \ -19.5 (\pm 3.3) $,
where we used the statistical package ASURV, which is developed
for the analysis of censored data
(ftp://www.astro.psu.edu/users/edf/asurv.shar).
\label{fig5}}
\end{center}
\end{figure}

\placefigure{fig5}

As was suggested by Saito (1997),
the slope in the 2--10 keV band is about 3/2 and is different from 
the slope of about one in the ROSAT band;
the spin-down effect appears in different ways for
different energy bands.
Furthermore, this energy-dependent effect seems common to
both populations.

In spite of 
a naive expectation that different magnetospheric parameters
in the two pulsar populations will result in a difference in the
X-ray radiation,
it is difficult to discriminate between the two populations
based on the $L_{\rm x}$-$\dot{E}$ plane.

A distinction between the millisecond pulsars
and the ordinary pulsars may be found in the pulsed emission.
Among pulsars with spin-down power ranging $10^{35} - 10^{37}$erg s$^{-1}$,
the six data points in Figure~5 for the {\it ordinary pulsars}
are all attributed to unpulsed emission,
while the three millisecond pulsars
(PSR~1937$+$21, PSR~1821--24 and PSR~J0218$+$4232) in the same range 
of spin-down power show pulse-dominated emission.
For the ordinary pulsars in Figure~5,
the Crab-like pulse-dominated pulsars are only found for
spin-down power larger than  $\sim 10^{37}$ erg s$^{-1}$, and the
pulsars in the range $\dot{E} \sim 10^{35}-10^{37}$ erg s$^{-1}$
are Vela-like, i.e., the emission is dominated by a DC component,
and only upper limits of the pulsed fraction are given, except for
the Vela pulsar itself.
RXTE observations of the Vela pulsar give a pulsed luminosity
of $\sim 5 \times 10^{30}$ erg s$^{-1}$
(\cite{shd99}),
which is three orders of magnitude smaller than the total X-ray luminosity
of $1.2 \times 10^{33}$ erg s$^{-1}$ (\cite{sai97})
plotted in Figure~5.
On the other hand, 
{PSR~1937$+$21}, {PSR~1821--24} and {PSR~J0218$+$4232} 
are in the range $\dot{E} \sim 10^{35}-10^{37}$ erg s$^{-1}$,
but their emission is dominated by the non-thermal pulsed emission.
If the {\it pulsed} luminosity can be measured for some of the ordinary
pulsars in the range $\dot{E} \sim 10^{35}-10^{37}$ erg s$^{-1}$, then
we may find that the two populations can be distinguished by their
location in the plane of {\it pulsed luminosity} versus rotation power.

Eight millisecond pulsars with still smaller spin-down power,
$\dot{E} \lsim 10^{35}$ erg s$^{-1}$, have been detected with ROSAT
(for PSR~J0030$+$0451, \cite{bek00}; for others, 
\cite{bkt99}).
Two of them, PSR~J0437--4715 and PSR~J2124--3358, are also detected with
ASCA (\cite{kts98}, \cite{sak99}).
Pulsation is detected for three pulsars:
PSR~J0437--4715 shows a sinusoidal shape and a pulsed fraction of about 30\%;
PSR~J2124--3358 shows a  double-peaked profile  and a pulsed 
fraction of $\sim 33$\%;
PSR~J0030$+$0451 also shows a double-peaked profile and a pulsed 
fraction of $\sim 69$\%.
The X-ray and radio profiles are similar to each other
for the latter two,
but profile alignment has not yet been done.
For ordinary pulsars with $\dot{E} \lsim 10^{35}$ erg s$^{-1}$, the radiation
is a mixture of several components including radiation from the cooling neutron
star, the heated polar caps, the nebula and the magnetosphere.
Decomposition of these components is not complete at present, 
and it is still difficult to compare the two pulsar populations 
for $\dot{E} \lsim 10^{35}$ erg s$^{-1}$.
Future and on-going missions should provide fruitful data to resolve the
problem.

A twin of PSR~1937$+$21 is PSR~1821--24, 
which has a similar rotation power and light cylinder field as PSR~1937$+$21
(see Table~2).
This pulsar also shows non-thermal narrow pulses with 
a double peaked profile in X-rays 
(\cite{skk97}, \cite{rot98}).
Pulse alignment for this pulsar shows that
the X-ray primary peak is nearly coincident with
the radio primary pulse.
It is known that
pulse alignment is exceptionally good for the Crab pulsar,
but is in general complicated for other ordinary pulsars,
(e.g., \cite{shd99} for the Vela pulsar;
\cite{rot98} for PSR~B1509-58; for other pulsars
see references therein).
It is interesting to note that the common feature among
these two millisecond pulsars and the Crab pulsar
is the light cylinder field strength of $\sim 10^6$ G
(Table~2). PSR~1937$+$21 shares another property, giant
radio pulses, with the Crab. Giant pulses of PSR~1937$+$21 are
delayed by some 40--50 $\mu$s from the main and interpulses
(\cite{cog96}), so are different in phase from
the X-ray pulses.

Another pulsar with high light-cylinder field ($B_L  = 3.2 \times 10^5$ G)
and with high spin-down power ($\dot{E} = 2.5 \times 10^{35}$ erg s$^{-1}$)
is PSR~J0218$+$4232, which shows non-thermal double peaked emission
(\cite{kui98}, \cite{min00}).
X-ray pulse alignment was not possible for this pulsar 
given the uncertainties in
absolute time of the ROSAT and BeppoSAX clocks.

The Crab-like features seen in the millisecond pulsars with
large light-cylinder fields may suggest that the 
X-ray and radio emission originates near the light cylinder.
However, the polar cap model can also account for the X-ray 
emission from the millisecond pulsar (\cite{luo00}).
Theoretical assessment of this issue is strongly suggested.

\acknowledgments
{\em Acknowledgments.} 
We are deeply indebted to K. Hashimotodani 
for his help in the spatial analysis,
to K. Ebisawa (GSFC) for valuable discussions about the timing analysis,
and to S. Kasahara for preparing figures for the X-ray images.
The Arecibo Observatory is part of the National Astronomy and
Ionosphere Center, which is operated by Cornell University under 
a cooperative agreement with the National Science Foundation.

This work is supported in part by a Grant-in-Aid for Scientific Research
(10117203, 12640229)
from the Ministry of Education, Science, and Culture in Japan.

\newpage
\begin{table}
\caption{Radio Ephemeris}
\begin{center}
\begin{tabular}{ll} \hline 
PSR name & 1937+21 \\
R.A.(J2000.0)       & 19 39 38.560 \\
Decl.(J2000.0)      & 21 34 59.14 \\
MJD validity interval & 50710 -- 50775 \\
$t_{0, geo}$        & 50742.000000003 \\
$\nu_{0}$           & 641.9282504747918 \\
$\dot{\nu}_{0}$     & $-4.33167 \times 10^{-14}$  \\
$\ddot{\nu}_{0}$    & 0.00  \\ \hline
\end{tabular} 
\par
Note --- The JPL DE200 solar system ephemeris file has been used.
The value of $t_{0, geo}$ is the  infinite-frequency 
geocentric UTC arrival time of the radio pulse (MJD), and
the integer part of $t_{0, geo}$ is the barycentric (TDB) epoch
of the pulsar position, pulse frequency and its time derivatives.
The ephemeris contains local values of the spin frequency and
dispersion measure; the other parameters were held fixed at the values
given in Kaspi et al. (1994).

\end{center}
\end{table}

\begin{table}
\caption{Parameters of the millisecond pulsars with 
large rotation power and the Crab pulsar}
\begin{tabular}{llllll}
	 \hline
         & $P$ & $\dot{P}$ & $\dot{E}$ & $B_{\rm s}$ & $B_{\rm L}$ \\
PSR name & ( $10^{-3}$ s ) & ( s s$^{-1}$ ) & ( erg s$^{-1}$ ) &
( G ) & ( G ) \\ \hline \hline
PSR B1937+21 &
 1.56        &
$1.05 \times 10^{-20}$&
$1.1  \times 10^{36}$ &
$8.2  \times 10^8$ & 
$1.0  \times 10^6$  
\\
PSR B1821--24 &
3.05 &
$1.62 \times 10^{-18}$ &
$2.2  \times 10^{36}$  &  
$4.5  \times 10^9$ &
$7.3  \times 10^5$ 
\\
PSR J0218+4232  &
2.32 &
$8.0  \times 10^{-20}$&
$2.5  \times 10^{35}$&
$8.7  \times 10^8  $&
$3.2  \times 10^5  $
\\
Crab  &
33.1 &
$4.22 \times 10^{-13}$ &
$4.6  \times 10^{38}$ &
$7.5  \times 10^{12}$ &
$0.9  \times 10^6 $
\\ \hline
\end{tabular}
\end{table}

\end{document}